\documentclass[aps,prb,reprint,amsmath,amssymb,floatfix,showkeys,superscriptaddress]{revtex4-2}
\makeatletter
\let\NAT@sort\z@
\makeatother
\usepackage[T1]{fontenc}
\usepackage{graphicx}
\usepackage{xcolor}
\definecolor{crossrefblue}{RGB}{0,0,139}
\usepackage[colorlinks=true,linkcolor=crossrefblue,citecolor=black,urlcolor=black]{hyperref}
\usepackage{placeins}
\newcommand{\modelnum}[1]{\uppercase\expandafter{\romannumeral#1\relax}}
\newcounter{figurepanel}[figure]

\begin{document}
\raggedbottom
\title{Scaling of phase rigidity without eigenstate coalescence}
\author{Shang-En Xie}
\affiliation{School of Physics, State Key Laboratory of Optoelectronic Materials and Technologies, and Guangdong Provincial Key Laboratory of Magnetoelectric Physics and Devices, Sun Yat-sen University, Guangzhou, China}

\author{Li-Mei Chen}
\affiliation{Key Laboratory of Atomic and Subatomic Structure and Quantum Control (Ministry of Education),
Guangdong Basic Research Center of Excellence for Structure and Fundamental Interactions of Matter,School of Physics, South China Normal University, Guangzhou 510006, China}
\affiliation{Guangdong Provincial Key Laboratory of Quantum Engineering and Quantum Materials,\mbox{Guangdong-Hong Kong Joint Laboratory of Quantum Matter, South China Normal University, Guangzhou 510006, China}}

\author{Yao Zhou}
\email{zhouyaode1@gmail.com}
\affiliation{School of Physics, State Key Laboratory of Optoelectronic Materials and Technologies, and Guangdong Provincial Key Laboratory of Magnetoelectric Physics and Devices, Sun Yat-sen University, Guangzhou, China}
\affiliation{Department of Physics and HK Institute of Quantum Science \& Technology, The University of Hong Kong, Hong Kong, China}

\author{Peng Ye}
\email{yepeng5@mail.sysu.edu.cn}
\affiliation{School of Physics, State Key Laboratory of Optoelectronic Materials and Technologies, and Guangdong Provincial Key Laboratory of Magnetoelectric Physics and Devices, Sun Yat-sen University, Guangzhou, China}
\date{\today}
\begin{abstract}
Phase rigidity vanishes as a power law near exceptional points, reflecting the coalescence of eigenstates at these points. In this Letter, we show that non-Hermitian skin effects induce distinct asymptotic behaviors of phase rigidity without eigenstate coalescence. To demonstrate this, we derive exact analytical results for three non-Hermitian chain models and verify them numerically. For the ordinary skin effect, phase rigidity decays exponentially with increasing system size $N$, as the right and left eigenstates are exponentially localized at opposite boundaries. For the scale-free skin effect, phase rigidity exhibits an $N^{-2}$ power-law approach to a nonzero constant as the right and left eigenstates have decay lengths proportional to $N$. For the algebraic skin effect studied here, phase rigidity can exhibit an $N^{-1/2}$ power-law decay to zero, with the right eigenstate algebraically localized near a boundary and the left eigenstate spread across the chain. These findings show that non-Hermitian skin effects govern phase-rigidity scaling by shaping the spatial profiles of the right and left eigenstates, enabling control of eigenstate nonorthogonality without exceptional points.
\end{abstract}
\maketitle

\textbf{\textit{Introduction}.}  
In non-Hermitian systems, phase rigidity quantifies eigenstate nonorthogonality through the overlap between left and right eigenstates~\cite{dingnonhermitian2022,wiersigpetermann2023,liunonhermitian2025,zhongobservation2026}.
The phase rigidity $r_j$ is defined as $|\langle L_j|R_j\rangle|/\sqrt{\langle L_j|L_j\rangle\langle R_j|R_j\rangle}$, where the right and left eigenstates satisfy $H\vert R_j\rangle=E_j\vert R_j\rangle$ and $\langle L_j\vert H=E_j\langle L_j\vert$, respectively.
Traditionally, phase rigidity is used to characterize exceptional points (EPs), near which it exhibits asymptotic power-law scaling~\cite{jinparitytimesymmetric2018,xiaoanisotropic2019,tangexceptional2020,zhanghighorder2020,jaiswalcharacterizing2023,chadhauncovering2024,caoobservation2024}. At an EP, the coalescing eigenstates are self-orthogonal, so $r_j=0$. Experimentally, phase rigidity also quantifies excess noise arising from eigenstate nonorthogonality through the Petermann factor $K_j=r_j^{-2}$~\cite{siegmanexcess1989,chengexperimental1996,wiersigpetermann2023,schomeruseigenvalue2024}. In non-Hermitian sensors, this relation helps determine whether enhanced responsivity translates into improved sensitivity in the presence of excess noise~\cite{wangpetermann2020,wiersigreview2020,schomerusnonreciprocal2020}.

For non-Hermitian lattice systems, an extensive number of bulk eigenstates can accumulate near boundaries, a phenomenon known as the non-Hermitian skin effect (NHSE)~\cite{yaoedge2018,leeanatomy2019,yokomizononbloch2019,okumatopological2020,zhanga2022,ammarimathematical2024}. For example, in the disorder-free Hatano--Nelson chain with open boundaries, the right and left eigenstates are exponentially localized at opposite ends with a size-independent decay length~\cite{kochbulkboundary2020,guoexact2021,edvardssonsensitivity2022}. For these states, the wave functions overlap less through their tails as the chain length $N$ increases. Phase rigidity is therefore expected to exhibit asymptotic scaling with $N$ in this chain, an observation that relates to the study of eigenvalue sensitivity in Ref.~\cite{noschese2013tridiagonal}. More generally, the NHSE takes several forms depending on the boundary conditions and lattice structure~\cite{liphasespace2025}. Specifically, in one-dimensional non-Hermitian systems, an impurity can induce the scale-free skin effect, where eigenstates accumulate near the boundaries and their decay length is proportional to the system size~\cite{liimpurity2021,moligninianomalous2023,wangobservation2025}. The critical NHSE exhibits a size-induced transition in which the eigenstate decay length changes from a fixed value in small systems to a size-dependent one in larger systems~\cite{licritical2020,yokomizoscaling2021}. The NHSE can also localize eigenstates at internal interfaces~\cite{weidemanntopological2020,hwangbulkedge2023,liuexperimental2023}, as well as at edges or corners of higher-dimensional systems~\cite{litopological2020,zhanguniversal2022}. In addition, the NHSE can give rise to wave functions that decay algebraically away from the boundary~\cite{zhangalgebraic2025,lialgebraic2026} or have Gaussian profiles~\cite{houdissolution2024}. Experimentally, the NHSE has been demonstrated in acoustic, photonic, electrical, and mechanical platforms, as well as ultracold atomic gases~\cite{xiaononhermitian2020,helbiggeneralized2020,hofmannreciprocal2020,zhangacoustic2021,zhangobservation2021,liunonhermitian2021,zouobservation2021,wangnonhermitian2022,liuexperimental2023,zhangelectrical2023,xiongtracking2024,wuspindependent2024,linobservation2024,sunphotonic2024,luextended2024,liobservation2024,huacoustic2025,wuhybridorder2025,wangcoupled2025,wangobservation2025,zhaotwodimensional2025,hanobservation2026}.

\begin{figure}[t]
\centering
\includegraphics[width=\columnwidth]{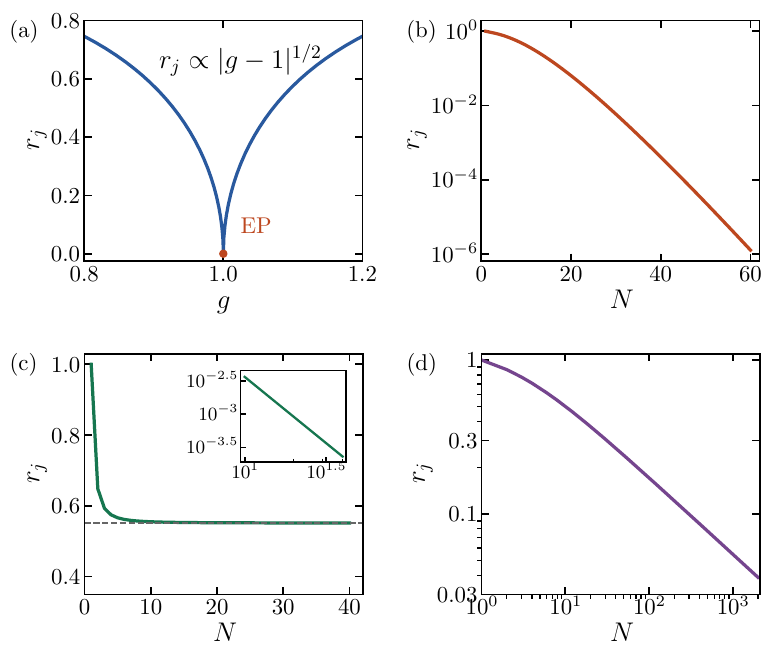}
\caption{Phase rigidity $r_j$ near an exceptional point and as a function of chain length $N$. (a) $r_j$ vanishes as $|g-1|^{1/2}$ near the EP at $g=1$ (red dot). (b) Exponential decay in the disorder-free Hatano--Nelson chain (Model~{\modelnum{1}}). (c) Power-law approach to a nonzero limit (dashed line) in the chain with a single impurity (Model~{\modelnum{2}}). The inset plots the deviation from this limit against chain length on logarithmic axes. (d) Power-law decay to zero in the chain with an imaginary gauge field (Model~{\modelnum{3}}).}
\label{fig:rigidity_ep_comparison}
\refstepcounter{figurepanel}\label{fig:rigidity_ep_comparison_ep}
\refstepcounter{figurepanel}\label{fig:rigidity_ep_comparison_ordinary}
\refstepcounter{figurepanel}\label{fig:rigidity_ep_comparison_impurity}
\refstepcounter{figurepanel}\label{fig:rigidity_ep_comparison_algebraic}
\end{figure}

The examples above show that eigenstates can accumulate at different locations and decay in different ways. The variety of non-Hermitian skin effects thus raises the question of how phase rigidity scales in these systems without EPs.

In this Letter, we show that the NHSE can induce distinct scaling behaviors of phase rigidity without exceptional points. Specifically, for the Hatano--Nelson chain without disorder, the right and left eigenstates are exponentially localized at opposite boundaries with a fixed decay length. By calculating their normalized overlap exactly, we find that phase rigidity decays exponentially at large $N$, in agreement with the numerical results and in contrast to its power-law scaling near an EP. For the non-Hermitian chain with a single impurity, our theoretical analysis shows that phase rigidity exhibits an $N^{-2}$ power-law approach to a nonzero limit. This limit arises because the left and right eigenstates accumulate on opposite sides of the impurity bond with decay lengths proportional to $N$. Finally, we consider an unusual algebraic skin effect that arises when a site-dependent imaginary gauge field is applied to a chain. The corresponding right eigenstate is algebraically localized near a boundary, while its left counterpart spreads across the chain. Using these eigenstate profiles, we show analytically that phase rigidity exhibits an $N^{-1/2}$ power-law decay to zero, as confirmed numerically. These findings reveal that the NHSE governs the scaling of phase rigidity through the spatial profiles of the left and right eigenstates, providing a mechanism for eigenstate nonorthogonality without coalescence.


\textbf{\textit{Phase rigidity near EPs}.}
We first consider the asymptotic behavior of phase rigidity near an EP in a two-level non-Hermitian model. Its Hamiltonian is
\begin{equation}
H_{\mathrm{EP}}(g)=
\begin{pmatrix}
0 & 1\\
g-1 & 0
\end{pmatrix},
\qquad g\in\mathbb{R}.
\end{equation}
The two eigenvalues are
\begin{equation}
E_{\pm}(g)=\pm\sqrt{g-1},
\end{equation}
with the corresponding right and left eigenstates proportional to $(1,E_\pm)^{\mathsf T}$ and $(E_\pm^*,1)^{\mathsf T}$, respectively, giving the phase rigidities
\begin{equation}
r_\pm(g)=\frac{2|E_\pm|}{1+|E_\pm|^2}
=\frac{2\sqrt{|g-1|}}{1+|g-1|}.
\end{equation}
Since $|E_+|=|E_-|$, the phase rigidities of the two eigenstates are equal, $r_+(g)=r_-(g)$. At $g=1$, they coalesce at a second-order EP and phase rigidity vanishes. On either side of the EP, $r_\pm(g)\propto |g-1|^{1/2}$ as $g$ approaches $1$, as shown in Fig.~\ref{fig:rigidity_ep_comparison_ep}.

For each of the three non-Hermitian chains below, we vary the chain length $N$ at fixed parameters to examine how different types of NHSE affect the scaling of phase rigidity.

\textbf{\textit{\texorpdfstring{Model~\modelnum{1}}{Model I}: Hatano--Nelson chain without disorder}.} 
\label{subsec:ordinarymodel}
We now turn to a disorder-free Hatano--Nelson chain~\cite{hatanolocalization1996,hatanovortex1997,hatanononhermitian1998} of $N$ sites with uniform nonreciprocal hopping and open boundary conditions (OBC). The Hamiltonian is
\begin{equation}
H_{\mathrm{\modelnum{1}}} = \sum_{n=1}^{N-1} \left( t_+ a_n^\dagger a_{n+1} + t_- a_{n+1}^\dagger a_n \right).
\label{eq:ordinaryH}
\end{equation}
Here $a_n^\dagger$ and $a_n$ are the creation and annihilation operators at site $n$, respectively. $t_+$ and $t_-$ are the positive leftward and rightward hopping amplitudes, respectively.

To evaluate phase rigidity in the lattice models considered here, we write it in the site basis as
\begin{equation}
r_j(N)=\frac{\left|\sum_{n=1}^{N} L_j^*(n) R_j(n)\right|}
{\sqrt{\left[\sum_{n=1}^{N}|R_j(n)|^2\right]
       \left[\sum_{n=1}^{N}|L_j(n)|^2\right]}},
\label{eq:rigidity}
\end{equation}
where $|\cdot|$ denotes the complex modulus. At site $n$, $R_j(n)=\langle n\vert R_j\rangle$ and $L_j(n)=\langle n\vert L_j\rangle$ are the complex amplitudes of the right and left eigenstates, respectively. To identify the eigenstates used in this expression, we select the eigenvalue $E_j$ with the largest imaginary part at each chain length $N$. If several eigenvalues share this maximum, we choose the one with the smallest real part.

After selecting $E_j$, we numerically solve the finite-chain transfer-matrix recurrences~\cite{molinaritransfer1998,kunstnonhermitian2019,yangexceptional2021,ghaemidizichehcompatibility2021,kawabatanonunitary2021,fuhybrid2023,luophotonic2023,kawabatahopf2025} to obtain the corresponding right and left eigenstates. We then compute $r_j(N)$ from these eigenstates using Eq.~\eqref{eq:rigidity}. For the Hatano--Nelson chain with $t_+=2$ and $t_-=1$, we plot the resulting phase rigidity as a function of $N$ in Fig.~\ref{fig:rigidity_ep_comparison_ordinary}. With $r_j$ on a logarithmic vertical axis and $N$ on a linear horizontal axis, the data follow an approximately straight line at large $N$, indicating that exponential decay dominates the scaling of phase rigidity with chain length. This exponential size dependence contrasts with the power-law scaling in $|g-1|$ near the EP.

\begin{figure}[!htb]
\centering
\includegraphics[width=\columnwidth]{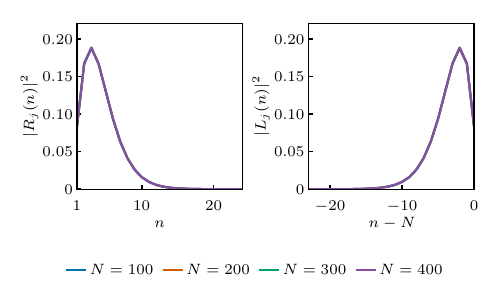}
\caption{Spatial profiles of the selected right and left eigenstates in the disorder-free Hatano--Nelson chain for $N=100,200,300,400$ and $(t_+,t_-)=(2,1)$. Each eigenstate has unit norm. The left panel shows $|R_j(n)|^2$ versus $n$, and the right panel shows $|L_j(n)|^2$ versus $n-N$. In each panel, the curves for different chain lengths nearly coincide over the boundary region shown.}
\label{fig:ordinaryskin}
\end{figure}

To understand this exponential decay, we plot the squared magnitudes $|R_j(n)|^2$ and $|L_j(n)|^2$ of the normalized right and left eigenstates (Fig.~\ref{fig:ordinaryskin}) and find that the two states are localized at opposite boundaries. These profiles suggest that both amplitudes have exponential envelopes with a size-independent decay length. We therefore infer that phase rigidity inherits its exponential dependence on chain length from the overlap of the wave-function tails.

\textbf{\textit{\texorpdfstring{Model~\modelnum{2}}{Model II}: Non-Hermitian chain with a single impurity}.}
\label{subsec:impuritymodel}
We next consider an $N$-site unidirectional chain with a single impurity. This chain exhibits the scale-free skin effect when the hopping amplitude on the impurity bond differs from the bulk hopping amplitude~\cite{moligninianomalous2023}. The Hamiltonian is
\begin{equation}
H_{\mathrm{\modelnum{2}}} = J\sum_{n=1}^{N-1} a_{n+1}^\dagger a_n + \gamma a_1^\dagger a_N.
\label{eq:impurityH}
\end{equation}
Here $a_n^\dagger$ and $a_n$ are the creation and annihilation operators at site $n$, respectively. $J$ is the positive bulk hopping amplitude from site $n$ to $n+1$, and $\gamma$ is the positive hopping amplitude from site $N$ to site $1$ across the impurity bond.

For this chain and the next, we select the eigenvalue and compute phase rigidity using the same procedure as for Model~{\modelnum{1}}. For $J=1$ and $\gamma=e^2$, we plot phase rigidity as a function of $N$ on linear axes in the main panel of Fig.~\ref{fig:rigidity_ep_comparison_impurity}. Phase rigidity exhibits a power-law approach to the nonzero limit marked by the dashed horizontal line. This saturation contrasts with the power-law decay near the EP and the exponential decay with chain length in Model~{\modelnum{1}}.

\begin{figure}[!htb]
\centering
\includegraphics[width=\columnwidth]{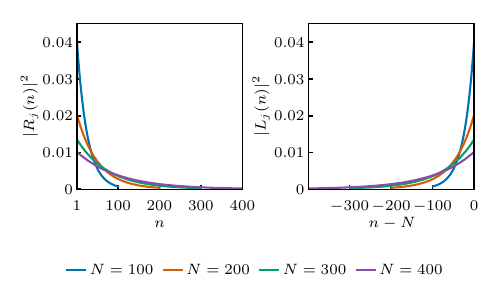}
\caption{Spatial profiles of the selected right and left eigenstates in the unidirectional chain with a single impurity for $N=100,200,300,400$ and $(J,\gamma)=(1,e^2)$. Each eigenstate has unit norm, and colors distinguish chain lengths. The left panel shows $|R_j(n)|^2$ versus the site index $n$, and the right panel shows $|L_j(n)|^2$ versus the signed displacement $n-N$ from site $N$. At fixed $N$, the right and left profiles are mirror images, peaking at sites $1$ and $N$, respectively. Both profiles broaden as $N$ increases, suggesting a decay length proportional to $N$.}
\label{fig:impurityskin}
\end{figure}

To understand this limiting behavior, we examine the spatial profiles in Fig.~\ref{fig:impurityskin}. The right and left eigenstates accumulate on opposite sides of the impurity bond, near sites $1$ and $N$, respectively. We observe that both profiles broaden as $N$ increases, suggesting that the decay length grows in proportion to $N$ and that each eigenstate extends over a fixed fraction of the chain. This broadening may account for the observed approach of phase rigidity to a nonzero limit. We will derive the limiting value and the leading finite-size correction shortly in this Letter.

\textbf{\textit{\texorpdfstring{Model~\modelnum{3}}{Model III}: Non-Hermitian chain with an imaginary gauge field}.}
We now examine phase-rigidity scaling in an $N$-site chain with OBC and a site-dependent imaginary gauge field, using the hopping Hamiltonian~\cite{longhierratic2025}
\begin{equation}
\mathcal{H}_{\mathrm{\modelnum{3}}}=\sum_n\left(t_n^{R}a_{n+1}^\dagger a_n+t_n^{L}a_n^\dagger a_{n+1}\right),
\end{equation}
where $t_n^{R}=t_0e^{h_n}$ and $t_n^{L}=t_0e^{-h_n} $ are the rightward and leftward hopping amplitudes, $t_0$ is the reciprocal hopping amplitude, and $h_n\in\mathbb{R}$ parameterizes a site-dependent imaginary gauge field that controls the local hopping asymmetry. Whereas Ref.~\cite{longhierratic2025} studies a random imaginary gauge field, we consider a deterministic field whose magnitude decreases with the site index. We choose $h_n=-\tfrac12\ln[(n+2)/n]$, set $t_0=-1$, and add an on-site potential of $-V$ at each endpoint. This gives
\begin{equation}
\begin{aligned}
H_{\mathrm{\modelnum{3}}}={}&-\sum_n
\left[
\sqrt{\frac{n}{n+2}}\,a_{n+1}^\dagger a_n
+\sqrt{\frac{n+2}{n}}\,a_n^\dagger a_{n+1}
\right]\\
&-V\left(a_1^\dagger a_1+a_N^\dagger a_N\right).
\end{aligned}
\label{eq:algebraicH}
\end{equation}
Here $a_n^\dagger$ and $a_n$ are the creation and annihilation operators at site $n$, respectively. For any real $V$, the Hamiltonian can be mapped to a Hermitian matrix by a diagonal similarity transformation. We set $V=1$ to simplify the spectral analysis. We compute the selected right and left eigenstates numerically and evaluate the corresponding phase rigidity using Eq.~\eqref{eq:rigidity}.

For this choice of $V$, we plot phase rigidity as a function of $N$ in Fig.~\ref{fig:rigidity_ep_comparison_algebraic}. With both axes on logarithmic scales, the data follow an approximately straight line at large $N$, indicating a power-law decay to zero. This behavior contrasts with the power-law approach to a nonzero limit in Model~{\modelnum{2}}.

\begin{figure}[!htb]
\centering
\includegraphics[width=\columnwidth]{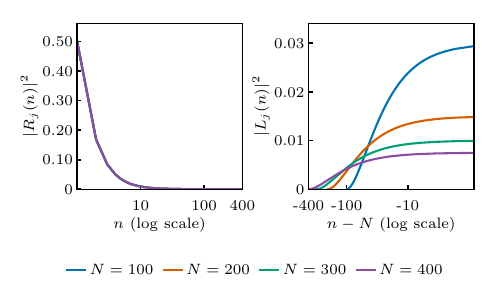}
\caption{Spatial profiles of the selected right and left eigenstates in the non-Hermitian chain with a site-dependent imaginary gauge field for $N=100,200,300,400$. Each eigenstate has unit norm, and colors distinguish chain lengths. The left panel shows $|R_j(n)|^2$ versus $\log_{10}n$, with ticks labeled by $n$. The right panel shows $|L_j(n)|^2$ versus $-\log_{10}(1+N-n)$, with ticks labeled by $n-N$. $|R_j(n)|^2$ decays algebraically away from the left boundary, while $|L_j(n)|^2$ rises toward the right boundary and extends over a finite fraction of the chain.}
\label{fig:algebraicprofiles}
\end{figure}

To understand this difference, we examine the spatial profiles in Fig.~\ref{fig:algebraicprofiles}. In Model~{\modelnum{2}}, the right and left eigenstates accumulate on opposite sides of the impurity bond. Here the selected right eigenstate is algebraically localized near the left boundary, while the corresponding left eigenstate extends over a substantial portion of the chain toward the right boundary. Despite these different profiles, a power law describes the deviation from the nonzero limit in Model~{\modelnum{2}} and the decay of phase rigidity to zero in Model~{\modelnum{3}}. We will shortly derive analytical expressions for the normalized overlap to determine how the details of the eigenstate profiles distinguish the two scaling behaviors.

\textbf{\textit{Analytic approach: Phase rigidity under the ordinary NHSE.}}
To explain the numerical results above, we evaluate the normalized overlap of the selected right and left eigenstates in each chain to obtain exact expressions for phase rigidity at finite $N$. We then expand these expressions at large $N$ and relate the distinct asymptotic behaviors of phase rigidity to the spatial profiles of the eigenstates.

In the Hatano--Nelson chain, the ordinary NHSE manifests as exponential localization of the right and left eigenstates at opposite boundaries, with a size-independent decay length. We derive the large-$N$ behavior of phase rigidity for the selected eigenstate of this chain and contrast it with the power-law decay near the EP in the two-level model. We start by solving for the right and left eigenstates.

Using the non-Bloch framework~\cite{yokomizononbloch2019}, we label the eigenstates by $m=1,\ldots,N$ and write the equation for the right eigenstate of the Hamiltonian in Eq.~\eqref{eq:ordinaryH} in the single-particle site basis as $E_{\mathrm{\modelnum{1}},m}R_{\mathrm{\modelnum{1}},m}(n)=t_-R_{\mathrm{\modelnum{1}},m}(n-1)+t_+R_{\mathrm{\modelnum{1}},m}(n+1)$. Substituting the bulk ansatz $R_{\mathrm{\modelnum{1}},m}(n)\propto\beta^n$ gives the characteristic equation
\begin{equation}
t_+\beta^2-E_{\mathrm{\modelnum{1}},m}\beta+t_-=0.
\end{equation}
The two spatial roots satisfy $\beta_1\beta_2=t_-/t_+$. For distinct roots, imposing the open boundary conditions $R_{\mathrm{\modelnum{1}},m}(0)=R_{\mathrm{\modelnum{1}},m}(N+1)=0$ on their linear combination gives $(\beta_1/\beta_2)^{N+1}=1$ (Appendix~\ref{app:ordinaryderivation}). The roots are therefore $\beta_{1,2}=\beta_{\mathrm{\modelnum{1}}}e^{\pm i m\pi/(N+1)}$, where the common root modulus is $\beta_{\mathrm{\modelnum{1}}}=\sqrt{t_-/t_+}$ and $\kappa_{\mathrm{\modelnum{1}}}=-\ln\beta_{\mathrm{\modelnum{1}}}=\tfrac12\ln(t_+/t_-)$. Substituting either root into $E_{\mathrm{\modelnum{1}},m}=t_-/\beta+t_+\beta$ gives the real eigenenergies
\begin{equation}
E_{\mathrm{\modelnum{1}},m}=2\sqrt{t_+t_-}\cos\frac{m\pi}{N+1}.
\end{equation}
Similarly, the left eigenstate obeys the recurrence $E_{\mathrm{\modelnum{1}},m}L_{\mathrm{\modelnum{1}},m}(n)=t_+L_{\mathrm{\modelnum{1}},m}(n-1)+t_-L_{\mathrm{\modelnum{1}},m}(n+1)$ with open boundary conditions $L_{\mathrm{\modelnum{1}},m}(0)=L_{\mathrm{\modelnum{1}},m}(N+1)=0$. We choose the initial amplitudes $R_{\mathrm{\modelnum{1}},m}(1)=L_{\mathrm{\modelnum{1}},m}(1)=\sqrt{2/(N+1)}\sin[m\pi/(N+1)]$, which fix the overall scales of the two eigenstates. Solving the two recurrences gives
\begin{equation}
\begin{aligned}
R_{\mathrm{\modelnum{1}},m}(n) &= \sqrt{\frac{2}{N+1}}\beta_{\mathrm{\modelnum{1}}}^{n-1}\sin\frac{m\pi n}{N+1},\\
L_{\mathrm{\modelnum{1}},m}(n) &= \sqrt{\frac{2}{N+1}}\beta_{\mathrm{\modelnum{1}}}^{-(n-1)}\sin\frac{m\pi n}{N+1}.
\end{aligned}
\label{eq:ordinaryselected}
\end{equation}

Using the eigenstate components in Eq.~\eqref{eq:ordinaryselected}, we obtain the squared norms of the right and left eigenstates,
\begin{equation}
\begin{aligned}
\langle R_{\mathrm{\modelnum{1}},m}\vert R_{\mathrm{\modelnum{1}},m}\rangle &= \frac{2}{N+1}\sum_{n=1}^{N}e^{-2\kappa_{\mathrm{\modelnum{1}}}(n-1)}\sin^2\!\frac{m\pi n}{N+1},\\
\langle L_{\mathrm{\modelnum{1}},m}\vert L_{\mathrm{\modelnum{1}},m}\rangle &= \frac{2}{N+1}\sum_{n=1}^{N}e^{2\kappa_{\mathrm{\modelnum{1}}}(n-1)}\sin^2\!\frac{m\pi n}{N+1}.
\end{aligned}
\end{equation}
For the amplitudes in Eq.~\eqref{eq:ordinaryselected}, the overlap evaluates to $\langle L_{\mathrm{\modelnum{1}},m}\vert R_{\mathrm{\modelnum{1}},m}\rangle=\frac{2}{N+1}\sum_{n=1}^{N}\sin^2[m\pi n/(N+1)]=1$.  Evaluating the resulting finite sum gives the exact phase rigidity for each eigenstate ($m=1,\ldots,N$):
\begin{equation}
r_m(N)=\frac{(N+1)\tanh\kappa_{\mathrm{\modelnum{1}}}}{\sinh[(N+1)\kappa_{\mathrm{\modelnum{1}}}]}
\left[1+\frac{\sinh^2\kappa_{\mathrm{\modelnum{1}}}}{\sin^2\theta_{\mathrm{\modelnum{1}},m}}\right],
\label{eq:ordinaryallmodes}
\end{equation}
where $\theta_{\mathrm{\modelnum{1}},m}=m\pi/(N+1)$.
The selected eigenstate of Model~\modelnum{1}  corresponds to $m=N$ at every chain length (Appendix~\ref{app:ordinaryderivation}).

At fixed $\kappa_{\mathrm{\modelnum{1}}}>0$, the factors in Eq.~\eqref{eq:ordinaryallmodes} at $m=N$ satisfy $\sinh[(N+1)\kappa_{\mathrm{\modelnum{1}}}]^{-1}=2e^{-(N+1)\kappa_{\mathrm{\modelnum{1}}}}[1+O(e^{-2(N+1)\kappa_{\mathrm{\modelnum{1}}}})]$ and $\sin^{-2}\theta_{\mathrm{\modelnum{1}},N}=(N+1)^2/\pi^2+O(1)$. These factors, together with the explicit $N+1$ in Eq.~\eqref{eq:ordinaryallmodes}, give a leading term proportional to $(N+1)^3e^{-\kappa_{\mathrm{\modelnum{1}}}N}$. Expanding $(N+1)^3=N^3[1+3/N+O(N^{-2})]$, we obtain
\begin{equation}
r_N(N) = C_{\mathrm{\modelnum{1}}} N^3 e^{-\kappa_{\mathrm{\modelnum{1}}} N} \left[1 + \frac{3}{N} + O(N^{-2})\right],
\label{eq:ordinaryasym}
\end{equation}
with the leading coefficient:
\begin{equation}
C_{\mathrm{\modelnum{1}}} = \frac{e^{\kappa_{\mathrm{\modelnum{1}}}}(1-e^{-2\kappa_{\mathrm{\modelnum{1}}}})^3}{2\pi^2(1+e^{-2\kappa_{\mathrm{\modelnum{1}}}})}.
\end{equation}

\begin{figure}[!htb]
\centering
\includegraphics[width=\columnwidth]{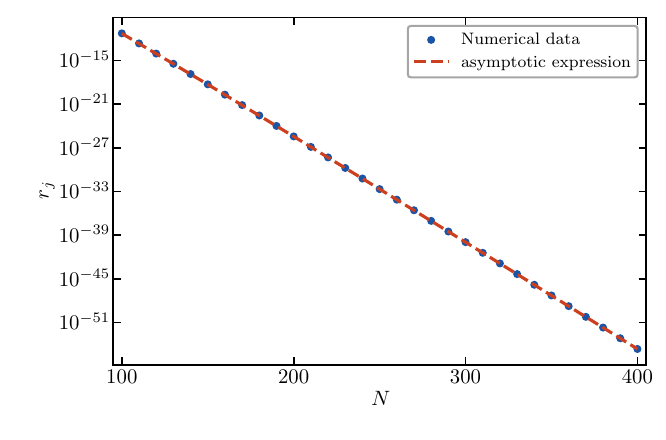}
\caption{Phase rigidity in the disorder-free Hatano--Nelson chain for $100\leq N\leq400$ in steps of $10$. The $N$ axis is linear and the $r_j$ axis is logarithmic (base 10). Blue points show phase rigidity calculated from right and left eigenstates obtained numerically using finite-chain transfer-matrix recurrences. The orange dashed curve shows the asymptotic form $C_{\mathrm{\modelnum{1}}}N^3e^{-\kappa_{\mathrm{\modelnum{1}}}N}$ with theoretical coefficients.}
\label{fig:windows_ordinary}
\end{figure}

We compute the phase rigidity shown in Fig.~\ref{fig:windows_ordinary} using the eigenstates obtained from the finite-chain boundary recurrences. In contrast to its power-law scaling near an EP, phase rigidity here approaches the asymptotic form $C_{\mathrm{\modelnum{1}}}N^3e^{-\kappa_{\mathrm{\modelnum{1}}}N}$ as $N$ increases. For the chosen parameters $t_+=2$ and $t_-=1$, we have $\kappa_{\mathrm{\modelnum{1}}}=(\ln2)/2$ and $C_{\mathrm{\modelnum{1}}}=\sqrt{2}/(24\pi^2)$. The large-$N$ asymptotic expression uses $\kappa_{\mathrm{\modelnum{1}}}\approx0.346574$ and $C_{\mathrm{\modelnum{1}}}\approx0.00597041$. The finite-size fitting analysis for this model is presented in Appendix~\ref{app:fitpredictions}.

\textbf{\textit{Analytic approach: Phase rigidity under the scale-free skin effect.}}
As discussed above, the spatial profiles of the right and left eigenstates broaden as $N$ increases. The decay length grows in proportion to the chain length. We call this the scale-free skin effect~\cite{liimpurity2021,moligninianomalous2023}. We compare the dependence of phase rigidity on chain length in this case with that for the ordinary NHSE.

To evaluate phase rigidity, we first obtain the right and left eigenstates. Applying the Hamiltonian in Eq.~\eqref{eq:impurityH} to a right eigenstate indexed by $m=1,\ldots,N$ gives the single-particle eigenvalue equations
\begin{equation}
\begin{aligned}
J R_{\mathrm{\modelnum{2}},m}(n-1) &= E_{\mathrm{\modelnum{2}},m}R_{\mathrm{\modelnum{2}},m}(n), && 2 \leq n \leq N,\\
\gamma R_{\mathrm{\modelnum{2}},m}(N) &= E_{\mathrm{\modelnum{2}},m}R_{\mathrm{\modelnum{2}},m}(1). &&
\end{aligned}
\end{equation}
We set $R_{\mathrm{\modelnum{2}},m}(1)=1$ and solve the bulk recurrence to obtain $R_{\mathrm{\modelnum{2}},m}(n)=\beta_{\mathrm{\modelnum{2}},m}^{n-1}$ with $\beta_{\mathrm{\modelnum{2}},m}=J/E_{\mathrm{\modelnum{2}},m}$. At site $1$, the boundary equation is $E_{\mathrm{\modelnum{2}},m}R_{\mathrm{\modelnum{2}},m}(1)=\gamma R_{\mathrm{\modelnum{2}},m}(N)$. Substitution gives $\gamma\beta_{\mathrm{\modelnum{2}},m}^{N-1}=J/\beta_{\mathrm{\modelnum{2}},m}$, hence $\beta_{\mathrm{\modelnum{2}},m}^{N}=J/\gamma$.

The corresponding left eigenstate satisfies $J L_{\mathrm{\modelnum{2}},m}(n+1)=E_{\mathrm{\modelnum{2}},m}^*L_{\mathrm{\modelnum{2}},m}(n)$ for $1\leq n<N$ and $\gamma L_{\mathrm{\modelnum{2}},m}(1)=E_{\mathrm{\modelnum{2}},m}^*L_{\mathrm{\modelnum{2}},m}(N)$. Similarly, taking $L_{\mathrm{\modelnum{2}},m}(1)=1$ gives $L_{\mathrm{\modelnum{2}},m}(n)=(\beta_{\mathrm{\modelnum{2}},m}^*)^{-(n-1)}$.

The exact energies and spatial roots are
\begin{equation}
\begin{aligned}
E_{\mathrm{\modelnum{2}},m} &= J(\gamma/J)^{1/N}e^{2\pi i m/N},\\
\beta_{\mathrm{\modelnum{2}},m} &= \frac{J}{E_{\mathrm{\modelnum{2}},m}} = (J/\gamma)^{1/N}e^{-2\pi i m/N}.
\label{eq:impurityroots}
\end{aligned}
\end{equation}
Eq.~\eqref{eq:impurityroots} shows that $|\beta_{\mathrm{\modelnum{2}},m}|=(J/\gamma)^{1/N}$ is independent of the eigenstate index $m$. We therefore define
\begin{equation}
\kappa_{\mathrm{\modelnum{2}}}(N) = -\ln|\beta_{\mathrm{\modelnum{2}},m}|=-\frac{\ln(J/\gamma)}{N}.
\end{equation}
For $\gamma\ne J$, the sign of $\kappa_{\mathrm{\modelnum{2}}}(N)$ determines the direction of decay of the right eigenstates, and its magnitude gives the inverse decay length.

For $\gamma=J$, $|\beta_{\mathrm{\modelnum{2}},m}|=1$ and $\kappa_{\mathrm{\modelnum{2}}}=0$. The recurrence relations give component magnitudes that are independent of $n$ for both eigenstates. The eigenstates are extended plane waves with real momenta in the conventional Brillouin zone, so the NHSE is absent. With the chosen initial amplitudes, $R_{\mathrm{\modelnum{2}},m}(n)=L_{\mathrm{\modelnum{2}},m}(n)$ at every site, so $r_m(N)=1$ at every size.

To study the scale-free skin effect, we focus on $\gamma\ne J$. The eigenstate amplitudes obtained from the transfer-matrix recurrences give
\begin{align}
\langle L_{\mathrm{\modelnum{2}},m}\vert R_{\mathrm{\modelnum{2}},m}\rangle&=\sum_{n=1}^{N}\beta_{\mathrm{\modelnum{2}},m}^{-(n-1)}\beta_{\mathrm{\modelnum{2}},m}^{n-1}=N,\notag\displaybreak[3]\\
\langle R_{\mathrm{\modelnum{2}},m}\vert R_{\mathrm{\modelnum{2}},m}\rangle&=\sum_{n=1}^{N}|\beta_{\mathrm{\modelnum{2}},m}|^{2(n-1)}
=\frac{1-(J/\gamma)^2}{1-(J/\gamma)^{2/N}}\notag\\*
&=(J/\gamma)^{(N-1)/N}
\frac{\sinh[N\kappa_{\mathrm{\modelnum{2}}}(N)]}{\sinh[\kappa_{\mathrm{\modelnum{2}}}(N)]},\notag\displaybreak[3]\\
\langle L_{\mathrm{\modelnum{2}},m}\vert L_{\mathrm{\modelnum{2}},m}\rangle&=\sum_{n=1}^{N}|\beta_{\mathrm{\modelnum{2}},m}|^{-2(n-1)}
=\frac{1-(\gamma/J)^2}{1-(\gamma/J)^{2/N}}\notag\\*
&=(\gamma/J)^{(N-1)/N}
\frac{\sinh[N\kappa_{\mathrm{\modelnum{2}}}(N)]}{\sinh[\kappa_{\mathrm{\modelnum{2}}}(N)]}.
\end{align}

The normalized overlap is independent of $m$, so the phase rigidity of the selected eigenstate of Model~\modelnum{2}   is
\begin{equation}
r_j(N)=\frac{N\sinh[\kappa_{\mathrm{\modelnum{2}}}(N)]}{\sinh[N\kappa_{\mathrm{\modelnum{2}}}(N)]}.
\label{eq:impurityexact}
\end{equation}

Expanding Eq.~\eqref{eq:impurityexact} for large $N$ gives
\begin{equation}
r_j(N) = C_{\mathrm{\modelnum{2}}}+\frac{D_{\mathrm{\modelnum{2}}}}{N^2}+O(N^{-4}),
\label{eq:impurityasym}
\end{equation}
where the thermodynamic limit and leading correction coefficient are
\begin{equation}
C_{\mathrm{\modelnum{2}}}=\frac{\ln(J/\gamma)}{\sinh[\ln(J/\gamma)]},\qquad
D_{\mathrm{\modelnum{2}}}=\frac{[\ln(J/\gamma)]^3}{6\sinh[\ln(J/\gamma)]}.
\end{equation}

\begin{figure}[!htb]
\centering
\includegraphics[width=\columnwidth]{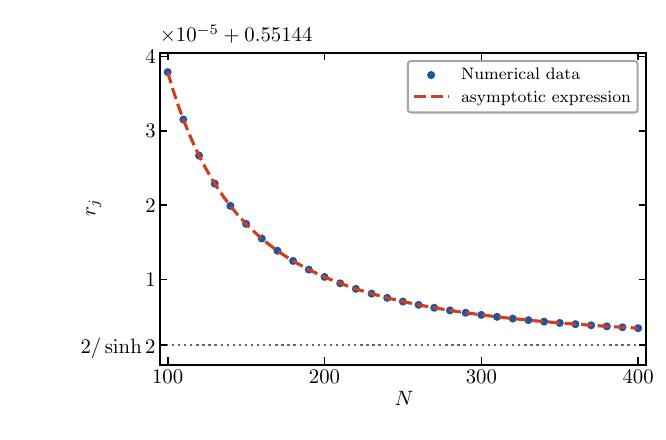}
\caption{Phase rigidity of the unidirectional chain with a single impurity for $100\leq N\leq400$ in steps of $10$. Both axes are linear, and the vertical axis uses an offset of $0.55144$ and a scale factor of $10^{-5}$. Blue points show phase rigidity from eigenstates computed numerically using transfer-matrix recurrences. The orange dashed curve shows the theoretical asymptotic form $C_{\mathrm{\modelnum{2}}}+D_{\mathrm{\modelnum{2}}}/N^2$ [Eq.~\eqref{eq:impurityasym}]. The gray dotted line marks the nonzero limit $C_{\mathrm{\modelnum{2}}}$, with the corresponding tick labeled $2/\sinh 2$.}
\label{fig:windows_impurity}
\end{figure}

For $J=1$ and $\gamma=e^2$, we test the analytical prediction by fitting the numerical results for phase rigidity to a constant plus a power-law correction, leaving the exponent free. The fit gives a limiting value consistent with $C_{\mathrm{\modelnum{2}}}$ and an exponent close to $2$ (Appendix~\ref{app:fitpredictions}). We also plot the theoretical curve $C_{\mathrm{\modelnum{2}}}+D_{\mathrm{\modelnum{2}}}/N^2$ alongside the numerical data in Fig.~\ref{fig:windows_impurity}, using $C_{\mathrm{\modelnum{2}}}=2/\sinh2\approx0.55144113$ and $D_{\mathrm{\modelnum{2}}}=4/(3\sinh2)\approx0.36762742$. The large-$N$ expansion gives a power-law approach to a nonzero constant in Model~{\modelnum{2}}, in sharp contrast to the exponential decay to zero in Model~{\modelnum{1}}.

\textbf{\textit{Analytic approach: Phase rigidity under the algebraic skin effect}.}  
As discussed above, the selected right eigenstate decays algebraically away from the boundary. We call this the algebraic skin effect and compare the resulting size dependence of phase rigidity with that for the ordinary NHSE and the scale-free skin effect.

For $V=1$, the on-site potential at each end of the chain in Eq.~\eqref{eq:algebraicH} is $-1$. The similarity transformation $H'_{\mathrm{\modelnum{3}}}=S^{-1}H_{\mathrm{\modelnum{3}}}S$, with $S$ diagonal and $S_{nn}=[n(n+1)]^{-1/2}$, maps the Hamiltonian to a Hermitian matrix $H'_{\mathrm{\modelnum{3}}}$. The eigenvalues and corresponding eigenstate components of $H'_{\mathrm{\modelnum{3}}}$ (Appendix~\ref{app:algebraicgroundstate}) are
\begin{equation}
\begin{aligned}
E_{\mathrm{\modelnum{3}},m}&=-2\cos\theta_{\mathrm{\modelnum{3}},m},\\
u_m(n)&=\cos[\theta_{\mathrm{\modelnum{3}},m}(n-\tfrac12)],
\end{aligned}
\end{equation}
where $\theta_{\mathrm{\modelnum{3}},m}=(N-m)\pi/N$ and $m=1,\ldots,N$.

The similarity transformation preserves the spectrum, so $H_{\mathrm{\modelnum{3}}}$ and $H'_{\mathrm{\modelnum{3}}}$ have the same eigenvalues. The corresponding right and left eigenstates of $H_{\mathrm{\modelnum{3}}}$ are obtained from $u_m$ as $S u_m$ and $S^{-1}u_m$, respectively. Their components are
\begin{equation}
\begin{aligned}
R_{\mathrm{\modelnum{3}},m}(n)
&=\frac{\cos[\theta_{\mathrm{\modelnum{3}},m}(n-\tfrac12)]}{\sqrt{n(n+1)}},\\
L_{\mathrm{\modelnum{3}},m}(n)
&=\sqrt{n(n+1)}\,\cos[\theta_{\mathrm{\modelnum{3}},m}(n-\tfrac12)].
\end{aligned}
\end{equation}

The spatial profiles can differ across eigenstates even though the hopping asymmetry is the same. For bonds $1\leq n<N$ with nonzero amplitudes at both sites, the ratio of adjacent amplitudes of the right eigenstate and the corresponding local decay rate are
\begin{equation}
\begin{aligned}
\beta_{\mathrm{\modelnum{3}},m}(n)
&=\frac{R_{\mathrm{\modelnum{3}},m}(n+1)}{R_{\mathrm{\modelnum{3}},m}(n)}\\
&=\sqrt{\frac{n}{n+2}}\,
  \frac{\cos[\theta_{\mathrm{\modelnum{3}},m}(n+\tfrac12)]}{\cos[\theta_{\mathrm{\modelnum{3}},m}(n-\tfrac12)]},\\
\kappa_{\mathrm{\modelnum{3}},m}(n)
&=-\ln|\beta_{\mathrm{\modelnum{3}},m}(n)|\\
&=\frac12\ln\!\left(\frac{n+2}{n}\right)
  +\ln\!\left|
    \frac{\cos[\theta_{\mathrm{\modelnum{3}},m}(n-\tfrac12)]}
         {\cos[\theta_{\mathrm{\modelnum{3}},m}(n+\tfrac12)]}
  \right|.
\end{aligned}
\end{equation}
For the left eigenstate, the ratio of the amplitudes at sites $n+1$ and $n$ is
\[
\frac{L_{\mathrm{\modelnum{3}},m}(n+1)}{L_{\mathrm{\modelnum{3}},m}(n)}
=\sqrt{\frac{n+2}{n}}\,
 \frac{\cos[\theta_{\mathrm{\modelnum{3}},m}(n+\tfrac12)]}
      {\cos[\theta_{\mathrm{\modelnum{3}},m}(n-\tfrac12)]}.
\]

Here, the selected eigenstate of Model~\modelnum{3}   also corresponds to $m=N$, with the lowest eigenvalue $E_{\mathrm{\modelnum{3}},N}=-2$ and $\theta_{\mathrm{\modelnum{3}},N}=0$. The spatial factor and decay rate reduce to $\beta_{\mathrm{\modelnum{3}},N}(n)=\sqrt{n/(n+2)}$ and $\kappa_{\mathrm{\modelnum{3}},N}(n)=\tfrac12\ln[(n+2)/n]$. The selected components are therefore
\begin{equation}
R_{\mathrm{\modelnum{3}},N}(n)=\frac{1}{\sqrt{n(n+1)}},\qquad
L_{\mathrm{\modelnum{3}},N}(n)=\sqrt{n(n+1)}.
\label{eq:algebraicstates}
\end{equation}

\begin{figure}[tbp]
\centering
\includegraphics[width=\columnwidth]{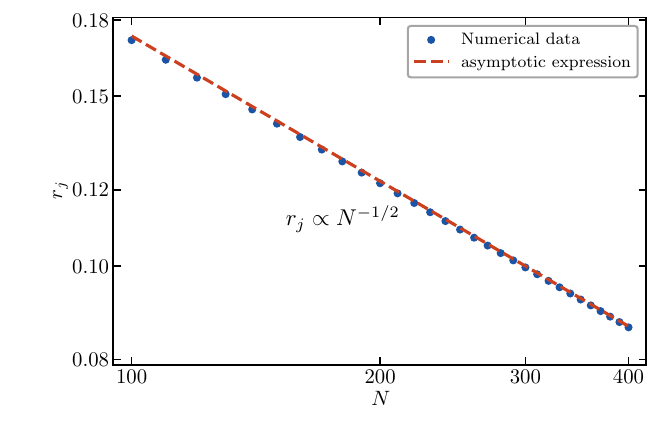}
\caption{Phase rigidity of the non-Hermitian chain with a site-dependent imaginary gauge field for $100\leq N\leq400$ in steps of $10$. Both axes use base-10 logarithmic scales. Blue points show phase rigidity from right and left eigenstates computed numerically using finite-chain transfer-matrix recurrences. The orange dashed curve shows the theoretical asymptotic form $\sqrt{3}\,N^{-1/2}$ [Eq.~\eqref{eq:algebraicasym}].}
\label{fig:algebraicfits}
\end{figure}

The sitewise product $L_{\mathrm{\modelnum{3}},N}(n)^*R_{\mathrm{\modelnum{3}},N}(n)$ in Eq.~\eqref{eq:algebraicstates} is unity, so the overlap is $N$. The squared norms are
\begin{equation}
\begin{aligned}
\langle R_{\mathrm{\modelnum{3}},N}\vert R_{\mathrm{\modelnum{3}},N}\rangle
&=\sum_{n=1}^N\left(\frac{1}{n}-\frac{1}{n+1}\right)
=\frac{N}{N+1},\\
\langle L_{\mathrm{\modelnum{3}},N}\vert L_{\mathrm{\modelnum{3}},N}\rangle
&=\sum_{n=1}^N(n^2+n)
=\frac{N(N+1)(N+2)}{3}.
\end{aligned}
\label{eq:algebraicnorms}
\end{equation}
Substitution into Eq.~\eqref{eq:rigidity} gives
\begin{equation}
r_N(N)=\sqrt{\frac{3}{N+2}}.
\label{eq:algebraicexact}
\end{equation}

Expanding Eq.~\eqref{eq:algebraicexact} at large $N$ gives
\begin{equation}
r_N(N)=\sqrt{3}\,N^{-1/2}
\left[1-\frac{1}{N}+O(N^{-2})\right].
\label{eq:algebraicasym}
\end{equation}
Together with $\langle L_{\mathrm{\modelnum{3}},N}\vert R_{\mathrm{\modelnum{3}},N}\rangle=N$, the norms in Eq.~\eqref{eq:algebraicnorms} give $r_N(N)\propto N^{-1/2}$.

As predicted by the large-$N$ expansion, we find numerically that the phase rigidity of the selected eigenstate in Model~{\modelnum{3}} approaches the asymptotic form $\sqrt{3}\,N^{-1/2}$ from below as $N$ increases from $100$ to $400$ (Fig.~\ref{fig:algebraicfits}). Model~{\modelnum{3}} thus exhibits a slow decay to zero, whereas in Model~{\modelnum{2}} the deviation from a nonzero limit decays more rapidly as $N^{-2}$. Appendix~\ref{app:fitpredictions} presents the finite-size fits for both models.

\textbf{\textit{Discussions}.}
We have shown that NHSEs can give rise to qualitatively distinct asymptotic behaviors of phase rigidity in the absence of EPs. Phase rigidity decays exponentially with $N$ for the ordinary NHSE, whereas it follows an $N^{-1/2}$ power law for the algebraic skin effect studied here. For the scale-free skin effect, by contrast, phase rigidity approaches a nonzero limit with an $N^{-2}$ correction. The NHSE thus offers an alternative way to control the asymptotic behavior of phase rigidity by shaping the spatial profiles of the right and left eigenstates.

A natural extension is to examine how the NHSE affects phase-rigidity scaling in higher-dimensional and multiband lattices. In higher dimensions, phase rigidity may scale differently for eigenstates that accumulate along edges or near corners. In multiband lattices, an open question is how internal orbital degrees of freedom and the different decay lengths of skin modes jointly affect the scaling.

For skin modes with a size-independent decay length, eigenstate nonorthogonality can affect driven responses, as illustrated by enhanced zero-mode response~\cite{budichnonhermitian2020,schomerusnonreciprocal2020} and nonlocal voltage amplification in topolectrical circuits~\cite{helbiggeneralized2020}. Eigenstate nonorthogonality also plays a role in boundary sensing~\cite{budichnonhermitian2020}, adiabatic amplification~\cite{ozawageometric2025}, and enhanced spectral sensitivity~\cite{nakai2024topological}. It remains to be determined how measurable signals scale with chain length in the scale-free and algebraic regimes studied here, where phase rigidity approaches a nonzero limit and decays as a power law, respectively. Given the exact relation between the conventional Petermann factor and phase rigidity, we ask whether an analogous relation holds for the generalized Petermann factor~\cite{zoudetecting2024}. A related question is whether the distinct phase-rigidity scalings found here are reflected in non-Hermitian entanglement, which is determined by the correlation matrix of the occupied eigenstates~\cite{chenquantum2022,chenquantum2024,zhou2026nonunitary}.

This work was in part supported by National Natural Science Foundation of China (NSFC) under Grants No.~12474149, Research Center for Magnetoelectric Physics of Guangdong Province under Grant No.~2024B0303390001, and Guangdong Provincial Key Laboratory of Magnetoelectric Physics and Devices under Grant No.~2026B1212170049.

\FloatBarrier
%

\appendix

\section{Derivation for the finite Hatano--Nelson chain without disorder}
\label{app:ordinaryderivation}
We start from the Hatano--Nelson Hamiltonian with open boundaries:
\begin{equation}
H_{\mathrm{\modelnum{1}}} = \sum_n \left( t_+ a_n^\dagger a_{n+1} + t_- a_{n+1}^\dagger a_n \right),
\end{equation}
We label the $N$ sites by $n=1,\ldots,N$. At site $n$, $a_n$ annihilates a particle and $a_n^\dagger$ creates one. The positive amplitudes $t_+$ and $t_-$ describe hopping to the left and right, respectively, with $t_+>t_-$.

Using the non-Bloch framework~\cite{yokomizononbloch2019}, we obtain the spatial roots from the bulk recurrence and impose OBC. In the single-particle subspace spanned by the site basis $\vert n\rangle = a_n^\dagger\vert0\rangle$, the bulk equation for eigenstate $m$ reads $E_{\mathrm{\modelnum{1}},m}R_{\mathrm{\modelnum{1}},m}(n)=t_-R_{\mathrm{\modelnum{1}},m}(n-1)+t_+R_{\mathrm{\modelnum{1}},m}(n+1)$, yielding the characteristic polynomial
\begin{equation}
t_+ \beta^2 - E_{\mathrm{\modelnum{1}},m}\beta + t_- = 0.
\end{equation}
At this eigenenergy, denote the two spatial roots by $\beta_1$ and $\beta_2$. Their product is independent of the eigenenergy, $\beta_1 \beta_2 = t_-/t_+$. In the generalized Brillouin zone (GBZ) construction, the bulk spectrum in the thermodynamic limit requires equimodular roots, leading to
\begin{equation}
\beta_{\mathrm{\modelnum{1}}} = \sqrt{t_-/t_+},\qquad
\kappa_{\mathrm{\modelnum{1}}} = -\ln\beta_{\mathrm{\modelnum{1}}} = \frac{1}{2}\ln(t_+/t_-).
\end{equation}
For distinct roots $\beta_1 \neq \beta_2$, the general bulk right eigenstate is a superposition of two spatial modes
\begin{equation}
R_{\mathrm{\modelnum{1}},m}(n) = A_m \beta_1^n + B_m \beta_2^n.
\end{equation}
Imposing OBC at the auxiliary sites $n=0$ and $n=N+1$ yields
\begin{widetext}
\begin{equation}
\begin{aligned}
R_{\mathrm{\modelnum{1}},m}(0) &= A_m + B_m = 0 \quad \Longrightarrow \quad B_m = -A_m,\\
R_{\mathrm{\modelnum{1}},m}(N+1) &= A_m(\beta_1^{N+1} - \beta_2^{N+1}) = 0 \quad \Longrightarrow \quad (\beta_1/\beta_2)^{N+1} = 1.
\end{aligned}
\end{equation}
\end{widetext}
The boundary constraint explicitly enforces $|\beta_1| = |\beta_2|$ at finite $N$. Combined with $\beta_1\beta_2 = \beta_{\mathrm{\modelnum{1}}}^2$, the allowed spatial roots are quantized as
\begin{equation}
\begin{aligned}
\beta_1 &= \beta_{\mathrm{\modelnum{1}}} e^{i\theta_{\mathrm{\modelnum{1}},m}},\\
\beta_2 &= \beta_{\mathrm{\modelnum{1}}} e^{-i\theta_{\mathrm{\modelnum{1}},m}},\\
\theta_{\mathrm{\modelnum{1}},m} &= \frac{m\pi}{N+1}, \quad m=1,\ldots,N.
\end{aligned}
\end{equation}
For the allowed indices $m=1,\ldots,N$, the quantized angles lie strictly between $0$ and $\pi$, and the two spatial roots are distinct. The corresponding right eigenstate is
\begin{equation}
R_{\mathrm{\modelnum{1}},m}(n) = 2iA_m \beta_{\mathrm{\modelnum{1}}}^n \sin(n\theta_{\mathrm{\modelnum{1}},m}) \propto \beta_{\mathrm{\modelnum{1}}}^{n-1} \sin(n\theta_{\mathrm{\modelnum{1}},m}).
\end{equation}
Substituting $\beta_1$ back into the dispersion $E_{\mathrm{\modelnum{1}},m} = t_-/\beta_1 + t_+\beta_1$ yields the purely real eigenenergies
\begin{equation}
E_{\mathrm{\modelnum{1}},m} = 2\sqrt{t_+t_-}\cos\theta_{\mathrm{\modelnum{1}},m}.
\label{eq:ordinaryenergies}
\end{equation}

For $t_+,t_->0$, Eq.~\eqref{eq:ordinaryenergies} gives $E_{\mathrm{\modelnum{1}},m}=2\sqrt{t_+t_-}\cos[m\pi/(N+1)]$, with $m=1,\ldots,N$. All eigenvalues are real, so they have the same imaginary part and the selection rule reduces to minimizing the real energy. The angles $m\pi/(N+1)$ increase strictly within $(0,\pi)$, where cosine decreases strictly. Explicitly, the cosine difference identity gives
\begin{equation}
\begin{aligned}
E_{\mathrm{\modelnum{1}},m+1}-E_{\mathrm{\modelnum{1}},m}
&=-4\sqrt{t_+t_-}\sin\frac{(2m+1)\pi}{2(N+1)}\\
&\quad\times\sin\frac{\pi}{2(N+1)}<0,
\\
& m=1,\ldots,N-1.
\end{aligned}
\end{equation}
Both sine factors are positive. Thus $E_{\mathrm{\modelnum{1}},1}>\cdots>E_{\mathrm{\modelnum{1}},N}$, and the rule of selecting the eigenvalue with the smallest real part uniquely picks $m=N$ at every size.

The corresponding left eigenstate satisfies $E_{\mathrm{\modelnum{1}},m}L_{\mathrm{\modelnum{1}},m}(n)=t_+L_{\mathrm{\modelnum{1}},m}(n-1)+t_-L_{\mathrm{\modelnum{1}},m}(n+1)$ with $L_{\mathrm{\modelnum{1}},m}(0)=L_{\mathrm{\modelnum{1}},m}(N+1)=0$. Replacing $\beta_{\mathrm{\modelnum{1}}}$ by $\beta_{\mathrm{\modelnum{1}}}^{-1}$ gives $L_{\mathrm{\modelnum{1}},m}(n)\propto\beta_{\mathrm{\modelnum{1}}}^{-(n-1)}\sin(n\theta_{\mathrm{\modelnum{1}},m})$. At the first site, we choose
\begin{equation}
R_{\mathrm{\modelnum{1}},m}(1)=L_{\mathrm{\modelnum{1}},m}(1)=\sqrt{\frac{2}{N+1}}\sin\theta_{\mathrm{\modelnum{1}},m}.
\end{equation}
The boundary equations then fix $R_{\mathrm{\modelnum{1}},m}(2)=E_{\mathrm{\modelnum{1}},m}R_{\mathrm{\modelnum{1}},m}(1)/t_+$ and $L_{\mathrm{\modelnum{1}},m}(2)=E_{\mathrm{\modelnum{1}},m}L_{\mathrm{\modelnum{1}},m}(1)/t_-$. These initial components determine all subsequent components through the bulk recurrences. The chosen first-site amplitudes fix only the overall scales, and the resulting eigenstates need not have unit norm. For the selected eigenstate $m=N$, this gives
\begin{equation}
\begin{aligned}
R_{\mathrm{\modelnum{1}},N}(n)&=\sqrt{\frac{2}{N+1}}\beta_{\mathrm{\modelnum{1}}}^{n-1}\sin\frac{N\pi n}{N+1},\\
L_{\mathrm{\modelnum{1}},N}(n)&=\sqrt{\frac{2}{N+1}}\beta_{\mathrm{\modelnum{1}}}^{-(n-1)}\sin\frac{N\pi n}{N+1}.
\end{aligned}
\end{equation}
The overlap of these states follows from the discrete sine identity:
\begin{equation}
\begin{aligned}
\sum_{n=1}^{N}\sin^2\frac{N\pi n}{N+1}&=\frac{N+1}{2},\\
\langle L_{\mathrm{\modelnum{1}},N}\vert R_{\mathrm{\modelnum{1}},N}\rangle&=\frac{2}{N+1}\sum_{n=1}^{N}\sin^2\frac{N\pi n}{N+1}=1.
\end{aligned}
\end{equation}

With $\beta_{\mathrm{\modelnum{1}}}=e^{-\kappa_{\mathrm{\modelnum{1}}}}$, the squared norms are
\begin{equation}
\begin{aligned}
\langle R_{\mathrm{\modelnum{1}},N}\vert R_{\mathrm{\modelnum{1}},N}\rangle&=\frac{2}{N+1}\sum_{n=1}^{N}e^{-2\kappa_{\mathrm{\modelnum{1}}}(n-1)}\sin^2\frac{N\pi n}{N+1},\\
\langle L_{\mathrm{\modelnum{1}},N}\vert L_{\mathrm{\modelnum{1}},N}\rangle&=\frac{2}{N+1}\sum_{n=1}^{N}e^{2\kappa_{\mathrm{\modelnum{1}}}(n-1)}\sin^2\frac{N\pi n}{N+1}.
\end{aligned}
\end{equation}
Changing the summation variable in the expression for $\|L_{\mathrm{\modelnum{1}},N}\|^2$ to $\ell=N+1-n$ and using $\sin^2[N\pi(N+1-\ell)/(N+1)]=\sin^2[N\pi\ell/(N+1)]$ gives
\begin{equation}
\begin{aligned}
\langle L_{\mathrm{\modelnum{1}},N}\vert L_{\mathrm{\modelnum{1}},N}\rangle&=\frac{2e^{2\kappa_{\mathrm{\modelnum{1}}}(N-1)}}{N+1}\times\sum_{\ell=1}^{N}e^{-2\kappa_{\mathrm{\modelnum{1}}}(\ell-1)}\sin^2\frac{N\pi\ell}{N+1}\\
&=e^{2\kappa_{\mathrm{\modelnum{1}}}(N-1)}\langle R_{\mathrm{\modelnum{1}},N}\vert R_{\mathrm{\modelnum{1}},N}\rangle.
\end{aligned}
\end{equation}
Substituting this relation and the calculated overlap into Eq.~\eqref{eq:rigidity} yields
\begin{equation}
r_N(N)=\frac{(N+1)e^{-\kappa_{\mathrm{\modelnum{1}}}(N-1)}}{2\displaystyle\sum_{n=1}^{N}e^{-2\kappa_{\mathrm{\modelnum{1}}}(n-1)}\sin^2\!\dfrac{N\pi n}{N+1}}.
\label{eq:ordinaryfinitesum_appendix}
\end{equation}
Its sine factor satisfies
\begin{equation}
\begin{aligned}
\sin\frac{N\pi n}{N+1}
&=\sin\left(n\pi-\frac{\pi n}{N+1}\right)\\
&=\sin(n\pi)\cos\frac{\pi n}{N+1}-\cos(n\pi)\sin\frac{\pi n}{N+1}\\
&=(-1)^{n+1}\sin\frac{\pi n}{N+1},\qquad n=1,\ldots,N.
\end{aligned}
\end{equation}
Thus, setting $\theta_{\mathrm{\modelnum{1}},1}=\pi/(N+1)$ gives $\sin^2[N\pi n/(N+1)]=\sin^2(n\theta_{\mathrm{\modelnum{1}},1})$. The sine factor alternates in sign and vanishes at the auxiliary sites $0$ and $N+1$.
Evaluating the geometric series via $\sin^2 x = (1-\cos 2x)/2$ yields the exact finite-sum identity
\begin{equation}
\begin{aligned}
&\sum_{n=1}^{N}e^{-2\kappa_{\mathrm{\modelnum{1}}}(n-1)}\sin^2(n\theta_{\mathrm{\modelnum{1}},1})\\
&\quad=\frac{(1+e^{-2\kappa_{\mathrm{\modelnum{1}}}})\mathcal{F}_{\mathrm{\modelnum{1}}}}
{(1-e^{-2\kappa_{\mathrm{\modelnum{1}}}})\mathcal{G}_{\mathrm{\modelnum{1}}}},\\
&\mathcal{F}_{\mathrm{\modelnum{1}}}=(1-e^{-2\kappa_{\mathrm{\modelnum{1}}}(N+1)})\sin^2\theta_{\mathrm{\modelnum{1}},1},\\
&\mathcal{G}_{\mathrm{\modelnum{1}}}=1-2e^{-2\kappa_{\mathrm{\modelnum{1}}}}\cos(2\theta_{\mathrm{\modelnum{1}},1})+e^{-4\kappa_{\mathrm{\modelnum{1}}}}.
\end{aligned}
\end{equation}
Substitution of this identity into Eq.~\eqref{eq:ordinaryfinitesum_appendix} and rewriting the exponential factors in hyperbolic form gives
\begin{equation}
r_N(N)=\frac{(N+1)\tanh\kappa_{\mathrm{\modelnum{1}}}}{\sinh[(N+1)\kappa_{\mathrm{\modelnum{1}}}]}
\left[1+\frac{\sinh^2\kappa_{\mathrm{\modelnum{1}}}}{\sin^2\theta_{\mathrm{\modelnum{1}},1}}\right],
\end{equation}
which is Eq.~\eqref{eq:ordinaryallmodes} evaluated at $m=N$. The same calculation applies to every eigenstate $m$: the identity $\sum_{n=1}^{N}\sin^2[m\pi n/(N+1)]=(N+1)/2$, the norm relation, and the finite-sum evaluation remain valid with the corresponding angle $\theta_{\mathrm{\modelnum{1}},m}=m\pi/(N+1)$, since $e^{2i(N+1)\theta_{\mathrm{\modelnum{1}},m}}=1$.

\section{Spectrum and eigenstates of the chain with a site-dependent imaginary gauge field}
\label{app:algebraicgroundstate}
We consider the non-Hermitian chain with a site-dependent imaginary gauge field at $V=1$, with Hamiltonian
\begin{equation}
\begin{aligned}
H_{\mathrm{\modelnum{3}}}={}&-\sum_n
\left[
\sqrt{\frac{n}{n+2}}\,a_{n+1}^\dagger a_n
+\sqrt{\frac{n+2}{n}}\,a_n^\dagger a_{n+1}
\right]\\
&-\left(a_1^\dagger a_1+a_N^\dagger a_N\right).
\end{aligned}
\end{equation}
Take the diagonal matrix $S$ with $S_{nn}=[n(n+1)]^{-1/2}$. Its entries are nonzero, so $S$ is invertible. Define the similarity-transformed Hamiltonian
\begin{equation}
H'_{\mathrm{\modelnum{3}}}=S^{-1}H_{\mathrm{\modelnum{3}}}S.
\end{equation}
Because $S$ is diagonal, each matrix element transforms as
\begin{equation}
(H'_{\mathrm{\modelnum{3}}})_{nm}=S_{nn}^{-1}(H_{\mathrm{\modelnum{3}}})_{nm}S_{mm}.
\end{equation}
For adjacent sites, the two hopping directions give
\begin{equation}
\begin{aligned}
(H'_{\mathrm{\modelnum{3}}})_{n,n+1}
&=\sqrt{n(n+1)}\left[-\sqrt{\frac{n+2}{n}}\right]\\
&\quad\times\frac{1}{\sqrt{(n+1)(n+2)}}=-1,\\
(H'_{\mathrm{\modelnum{3}}})_{n+1,n}
&=\sqrt{(n+1)(n+2)}\left[-\sqrt{\frac{n}{n+2}}\right]\\
&\quad\times\frac{1}{\sqrt{n(n+1)}}=-1.
\end{aligned}
\end{equation}
Thus the transformation removes the hopping asymmetry. For diagonal entries, the factors $S_{nn}^{-1}$ and $S_{nn}$ cancel, leaving the on-site potentials unchanged. For $N\geq2$, the nonzero entries are
\begin{equation}
\begin{aligned}
(H'_{\mathrm{\modelnum{3}}})_{nn}&=
\begin{cases}
-1,&n=1,N,\\
0,&1<n<N,
\end{cases}\\
(H'_{\mathrm{\modelnum{3}}})_{n,n+1}&=(H'_{\mathrm{\modelnum{3}}})_{n+1,n}=-1,\quad 1\leq n<N,
\end{aligned}
\end{equation}
and all other entries vanish. These real, symmetric matrix elements explicitly show that $H'_{\mathrm{\modelnum{3}}}=(H'_{\mathrm{\modelnum{3}}})^\dagger$.

The characteristic polynomial is $D_N(E)=\det(EI-H'_{\mathrm{\modelnum{3}}})$, where $I$ is the $N\times N$ identity. The nonzero entries of $EI-H'_{\mathrm{\modelnum{3}}}$ can be written as
\begin{equation}
\begin{aligned}
(EI-H'_{\mathrm{\modelnum{3}}})_{nn}
&=\begin{cases}
E+1,&n=1,N,\\
E,&1<n<N,
\end{cases}\\
(EI-H'_{\mathrm{\modelnum{3}}})_{n,n+1}
&=(EI-H'_{\mathrm{\modelnum{3}}})_{n+1,n}=1,
\end{aligned}
\end{equation}
where the second line holds for $1\leq n<N$.

The matrix $H''_{\mathrm{\modelnum{3}}}$ is obtained from $EI-H'_{\mathrm{\modelnum{3}}}$ in two steps.
\begin{enumerate}
\item Replace column $j$ by the sum of columns $j,j+1,\ldots,N$ of the original matrix. The first column contains the original row sums, all equal to $E+2$.
\item Starting from the bottom, subtract row $i-1$ from row $i$ for $i=N,N-1,\ldots,2$. The first row stays unchanged.
\end{enumerate}
The entries of $H''_{\mathrm{\modelnum{3}}}$ are
\begin{equation}
\begin{aligned}
(H''_{\mathrm{\modelnum{3}}})_{ii}
&=\begin{cases}
E+2,&i=1,\\
E,&2\leq i\leq N,
\end{cases}\\
(H''_{\mathrm{\modelnum{3}}})_{12}&=1,\\
(H''_{\mathrm{\modelnum{3}}})_{i,i+1}&=(H''_{\mathrm{\modelnum{3}}})_{i+1,i}=1,\quad 2\leq i<N,
\end{aligned}
\end{equation}
with all other entries zero. Thus the first column is $(E+2,0,\ldots,0)^{\mathsf T}$, and the lower-right block can be viewed as the tridiagonal matrix $T_{N-1}(E)$. Its entries are related to those of $H''_{\mathrm{\modelnum{3}}}(E)$ by
\begin{equation}
\begin{aligned}
[T_{N-1}(E)]_{ij}&=[H''_{\mathrm{\modelnum{3}}}(E)]_{i'j'}\\
&=\begin{cases}
E,&i=j,\\
1,&|i-j|=1,\\
0,&|i-j|>1,
\end{cases}
\end{aligned}
\end{equation}
where $1\leq i,j\leq N-1$, $i'=i+1$, and $j'=j+1$.

Both operations preserve the determinant, so $D_N(E)=\det(H''_{\mathrm{\modelnum{3}}})$. Expansion along the first column therefore gives
\begin{equation}
D_N(E)=(E+2)\det T_{N-1}(E).
\end{equation}
For a chain of length $N+1$, expanding the determinant of the corresponding block $T_N(E)$ along the last row gives $\det T_N=E\det T_{N-1}-\det T_{N-2}$. Multiplying by $E+2$ gives
\begin{equation}
D_{N+1}=ED_N-D_{N-1},\qquad N\geq3,
\end{equation}
with $D_2=E(E+2)$ and $D_3=(E+2)(E^2-1)$ from the one- and two-dimensional tridiagonal blocks.

Set $E=z+z^{-1}$, where $z\ne0$. This substitution covers every $E$, since $z^2-Ez+1=0$ always has nonzero complex roots. For $z\ne\pm1$ and $N\geq2$, the closed form is
\begin{equation}
D_N(z+z^{-1})=(E+2)\frac{z^N-z^{-N}}{z-z^{-1}}.
\label{eq:algebraiccharacteristic}
\end{equation}
Substituting Eq.~\eqref{eq:algebraiccharacteristic} at sizes $N$ and $N-1$ into the determinant recurrence gives the numerator
\begin{equation}
\begin{aligned}
&(z+z^{-1})(z^N-z^{-N})-(z^{N-1}-z^{1-N})\\
&\qquad=z^{N+1}-z^{-(N+1)}.
\end{aligned}
\end{equation}
This proves Eq.~\eqref{eq:algebraiccharacteristic} by induction.

At $z=\pm1$, the numerator and denominator of the quotient in Eq.~\eqref{eq:algebraiccharacteristic} both vanish. The finite geometric sum
\begin{equation}
\frac{z^N-z^{-N}}{z-z^{-1}}
=\sum_{\ell=0}^{N-1}z^{N-1-2\ell}
\end{equation}
extends the quotient to these two points. At $z=1$, the sum equals $N$, so $D_N(2)=4N\ne0$. Thus $E=2$ is not an eigenvalue. At $z=-1$, every exponent has the same parity as $N-1$, and the sum equals $(-1)^{N-1}N$. Hence $E=-2$ is a simple root from the factor $E+2$.

The remaining roots satisfy $z^{2N}=1$ with $z\ne\pm1$. Each pair $z,z^{-1}$ gives the same energy. Taking $z=\exp(ik\pi/N)$ for $k=1,\ldots,N-1$ yields the $N-1$ distinct energies $2\cos(k\pi/N)$. Including $E=-2$, the full spectrum is
\begin{equation}
E_{\mathrm{\modelnum{3}},m}=-2\cos\theta_{\mathrm{\modelnum{3}},m},
\end{equation}
where $\theta_{\mathrm{\modelnum{3}},m}=(N-m)\pi/N$ and $m=1,\ldots,N$. The similarity transformation gives the same spectrum for $H_{\mathrm{\modelnum{3}}}$.

We use a transfer matrix to compute the eigenstate components of $H'_{\mathrm{\modelnum{3}}}$. For the bulk sites $2\leq n\leq N-1$, the rows of the Hermitian matrix give
\begin{equation}
u(n+1)=2\cos\theta_{\mathrm{\modelnum{3}},m}\,u(n)-u(n-1).
\end{equation}
The first row gives the left boundary condition
\begin{equation}
u(2)=(2\cos\theta_{\mathrm{\modelnum{3}},m}-1)u(1).
\end{equation}

Write the bulk recurrence in transfer-matrix form,
\begin{equation}
       \begin{aligned}
       \begin{pmatrix}u(n+1)\\u(n)\end{pmatrix}
&=M_m\begin{pmatrix}u(n)\\u(n-1)\end{pmatrix},\\
M_m&=\begin{pmatrix}2\cos\theta_{\mathrm{\modelnum{3}},m}&-1\\1&0\end{pmatrix}.
       \end{aligned}
\end{equation}
For $0<\theta_{\mathrm{\modelnum{3}},m}<\pi$, the $k$th power is
\begin{equation}
\begin{aligned}
M_m^k &= \frac{1}{\sin\theta_{\mathrm{\modelnum{3}},m}}\\
&\quad{}\times
\begin{pmatrix}
\sin[(k+1)\theta_{\mathrm{\modelnum{3}},m}]&-\sin(k\theta_{\mathrm{\modelnum{3}},m})\\
\sin(k\theta_{\mathrm{\modelnum{3}},m})&-\sin[(k-1)\theta_{\mathrm{\modelnum{3}},m}]
\end{pmatrix}.
\end{aligned}
\end{equation}
At $k=0$, this is the identity. Left multiplication by $M_m$ and the sine addition formula give the same expression with $k$ replaced by $k+1$.

For $2\leq n\leq N$, applying $M_m^{n-2}$ to $(u(2),u(1))^{\mathsf T}$ gives $(u(n),u(n-1))^{\mathsf T}$. Taking the first row of this matrix power and using $u(2)=(2\cos\theta_{\mathrm{\modelnum{3}},m}-1)u(1)$ gives
\begin{equation}
\begin{aligned}
\frac{u(n)}{u(1)}
&=(2\cos\theta_{\mathrm{\modelnum{3}},m}-1)(M_m^{n-2})_{11}
  +(M_m^{n-2})_{12}\\
&=\frac{\sin(n\theta_{\mathrm{\modelnum{3}},m})-\sin[(n-1)\theta_{\mathrm{\modelnum{3}},m}]}{\sin\theta_{\mathrm{\modelnum{3}},m}}\\
&=\frac{\cos[(n-\tfrac12)\theta_{\mathrm{\modelnum{3}},m}]}{\cos(\theta_{\mathrm{\modelnum{3}},m}/2)}.
\end{aligned}
\end{equation}
Direct substitution confirms that these amplitudes satisfy the right boundary equation $(2\cos\theta_{\mathrm{\modelnum{3}},m}-1)u(N)-u(N-1)=0$. For $m=N$, $\theta_{\mathrm{\modelnum{3}},N}=0$ and the left boundary equation gives $u(2)=u(1)$. The bulk recurrence then gives a constant profile, which also satisfies the right boundary equation. Choosing the first-site amplitude $u(1)=\cos(\theta_{\mathrm{\modelnum{3}},m}/2)$ gives the Hermitian eigenstate components
\begin{equation}
u_m(n)=\cos[\theta_{\mathrm{\modelnum{3}},m}(n-\tfrac12)],\qquad 1\leq n\leq N.
\end{equation}

\section{Finite-size effect analysis}
\label{app:fitpredictions}
For each model, we fit the phase rigidity data over four size windows ($100$--$200$, $200$--$300$, $300$--$400$, and $100$--$400$), minimizing the sum of squared residuals in phase rigidity. For all three models, phase rigidity is computed from the numerical boundary solutions using the normalized-overlap definition. The fits use every integer $N$ within each size window. The rows labeled $N\to\infty$ give the theoretical parameter values of the leading large-$N$ forms.

\setcounter{table}{0}
\subsection{Hatano--Nelson chain without disorder}
To examine how finite-size sampling affects the extraction of scaling exponents, we fit the numerical phase rigidity data to the general ansatz
\begin{equation}
r_{\mathrm{fit}}(N) = a N^\alpha e^{-b N}.
\label{eq:ordinaryfit}
\end{equation}
We fit Eq.~\eqref{eq:ordinaryfit} separately over four size windows. As the fixed-width fitting window shifts to larger sizes, the fitted exponent $\alpha$ increases monotonically from $2.89989$ on the $100$--$200$ window to $2.97510$ on the $300$--$400$ window, approaching the theoretical value $\alpha=3$. Concurrently, the fitted decay rate $b$ approaches the analytical rate $\kappa_{\mathrm{\modelnum{1}}}$. The fitted parameters depend on the size window because the fitting ansatz omits the finite-size corrections in Eq.~\eqref{eq:ordinaryasym}.

The $100$--$400$ fit reproduces the $100$--$200$ parameters to the digits shown, so the finite-size bias in the fitted parameters persists even over the full size window.

Table~\ref{tab:ordinarywindows} lists the fitted parameters and their theoretical values in the large-$N$ limit.

\begin{table}[!htbp]
\caption{Hatano--Nelson chain without disorder: parameters fitted to the numerical phase rigidity values in each size window.}
\label{tab:ordinarywindows}
\fontsize{7.5}{9}\selectfont
\setlength{\tabcolsep}{2.5pt}
\begin{ruledtabular}
\begin{tabular}{rccc}
Window&$a$&$\alpha$&$b$\\
100--200&0.00930977452157&2.89989412389419&0.34602722862299\\
200--300&0.00737450346157&2.95922706266695&0.34646453544152\\
300--400&0.00686609323724&2.97510169443708&0.34652983941339\\
100--400&0.00930977452157&2.89989412389419&0.34602722862299\\
\hline
$N\to\infty$&0.005970408003&3&0.346573590280\\
\end{tabular}
\end{ruledtabular}
\end{table}

\FloatBarrier

\subsection{Unidirectional chain with a single impurity}
We fit the convergence toward the asymptotic limit using the three-parameter ansatz
\begin{equation}
r_{\mathrm{fit}}(N) = r_0 + d N^{-\eta},
\end{equation}
where the exponent $\eta$ is unconstrained. The fitted exponent $\eta$ decreases from $2.00004417$ in the $100$--$200$ window to $2.00000678$ in the $300$--$400$ window, approaching the analytical value $2$. The fitted coefficients $r_0$ and $d$ also approach the theoretical values $C_{\mathrm{\modelnum{2}}}$ and $D_{\mathrm{\modelnum{2}}}$, respectively.

The value $\eta=2.00002902$ obtained over the full size window remains farther from $2$ than the $300$--$400$ result, reflecting the influence of stronger finite-size corrections at small $N$.

Table~\ref{tab:impuritywindows} lists the fitted parameters and their theoretical values in the large-$N$ limit.

\begin{table}[!htbp]
\caption{Unidirectional chain with a single impurity: parameters fitted to the numerical phase rigidity values in each size window.}
\label{tab:impuritywindows}
\fontsize{7.5}{9}\selectfont
\setlength{\tabcolsep}{2.5pt}
\begin{ruledtabular}
\begin{tabular}{rccc}
Window&$r_0$&$d$&$\eta$\\
100--200&0.55144112974711&0.36770741345823&2.00004417242990\\
200--300&0.55144112956472&0.36765527827473&2.00001379711221\\
300--400&0.55144112954876&0.36764198985310&2.00000678262736\\
100--400&0.55144112961082&0.36768283041530&2.00002902457370\\
\hline
$N\to\infty$&0.551441129544&0.367627419696&2\\
\end{tabular}
\end{ruledtabular}
\end{table}

\FloatBarrier

\subsection{Non-Hermitian chain with a site-dependent imaginary gauge field}
We fit the numerical phase rigidity data for $100\leq N\leq400$ to
\begin{equation}
r_{\mathrm{fit}}(N)=A N^{-\alpha}.
\label{eq:algebraicfit}
\end{equation}
The fitted exponent $\alpha$ approaches $1/2$ as the fitting window moves to larger sizes. The finite-size shift $N+2$ in Eq.~\eqref{eq:algebraicexact} explains why the fitted exponent stays below $1/2$.

The $100$--$400$ exponent $\alpha=0.49480688$ lies farther below $1/2$ than the $300$--$400$ value, showing that finite-size effects from small $N$ persist in the fit over the full size window.

Table~\ref{tab:algebraicwindows} lists the fitted parameters and their theoretical large-$N$ values.

\begin{table}[!htbp]
\caption{Parameters of the power-law fits in Eq.~\eqref{eq:algebraicfit} for the non-Hermitian chain with a site-dependent imaginary gauge field.}
\label{tab:algebraicwindows}
\begin{ruledtabular}
\begin{tabular}{rcc}
Window & $A$ & $\alpha$ \\
100--200&1.66059357&0.49293824\\
200--300&1.68680567&0.49593251\\
300--400&1.69824589&0.49712332\\
100--400&1.67600135&0.49480688\\
\hline
$N\to\infty$ & $1.73205081$ & $0.5$ \\
\end{tabular}
\end{ruledtabular}
\end{table}

\FloatBarrier

\end{document}